\documentclass[11pt]{article}
\usepackage{amssymb,latexsym,amsmath}
\usepackage{amsfonts, graphics}
\newcommand{\R}{\mathbb R}

\def\open#1{\setbox0=\hbox{$#1$}
\baselineskip = 0pt
\vbox{\hbox{\hspace*{0.4 \wd0}\tiny $\circ$}\hbox{$#1$}}
\baselineskip = 11pt\!}
\def\fn{\open{f}}

\def\be{\begin{equation}}
\def\ee{\end{equation}}
\def\bea{\begin{eqnarray}}
\def\eea{\end{eqnarray}}
\def\beas{\begin{eqnarray*}}
\def\eeas{\end{eqnarray*}}

\def\supp{\mathrm{supp}\,}

\newtheorem{theorem}{Theorem}
\newtheorem{lemma}{Lemma}

\def\open#1{\setbox0=\hbox{$#1$}
\baselineskip = 0pt
\vbox{\hbox{\hspace*{0.4 \wd0}\tiny $\circ$}\hbox{$#1$}}
\baselineskip = 11pt\!}
\def\fn{\open{f}}
\def\supp{\mathrm{supp}\,}

\begin{document}
\title{Regularity results for the spherically symmetric Einstein-Vlasov system}
\author{H\aa kan Andr\'{e}asson\\
Mathematical Sciences\\ University of Gothenburg\\
        Mathematical Sciences\\ Chalmers University of Technology\\
        S-41296 G\"oteborg, Sweden\\
        email: hand@chalmers.se}
\maketitle

\begin{abstract}
The spherically symmetric Einstein-Vlasov system is considered in Schwarzschild coordinates and in
maximal-isotropic coordinates. An open problem is the issue of global existence for initial data
without size restrictions. The main purpose of the present work is to propose a method of approach for general initial data, which improves the regularity of the terms that need to be estimated compared to previous methods. We prove that global existence holds outside the centre in both these coordinate systems. In the Schwarzschild case we improve the bound on the momentum support obtained in \cite{RRS} for compact initial data. The improvement implies that we can admit non-compact data with both ingoing and outgoing matter. This extends one of the results in \cite{AR1}. In particular our method avoids the difficult task of treating the pointwise matter terms.
Furthermore, we show that singularities never form in Schwarzschild time for ingoing matter as long as $3m\leq r.$ This removes an additional assumption made in \cite{A1}. Our result in maximal-isotropic coordinates is analogous to the result in \cite{R1}, but our method is different and it improves the regularity of the terms that need to be estimated for proving global existence in general.
\end{abstract}

\section{Introduction}
\setcounter{equation}{0}
In the present work we investigate the issue of global existence for the spherically symmetric Einstein-Vlasov system when the initial data is unrestricted in size. The system is analyzed in Schwarzschild coordinates, i.e. in a polar time gauge, and in maximal-isotropic coordinates where a maximal time gauge is imposed. These coordinate systems are often, in the literature, conjectured to be singularity avoiding. However, there is to our knowledge no proof of this statement for any matter model and it would be very satisfying to provide an answer to this conjecture for the Einstein-Vlasov system. Moreover, a proof of global existence would be of great importance due to its relation to the weak cosmic censorship conjecture, cf. \cite{Cu1,Ds1,DR,A1,AKR2,AKR3}. A third motivation for our interest in this problem is the fact that global existence for general data have been obtained for the Vlasov-Poisson system, i.e. Newtonian gravity coupled to Vlasov matter. Batt \cite{B} showed global existence in the spherically symmetric case 1977, and the general case was settled independently by Pfaffelmoser \cite{P}, and Lions and Perthame \cite{LP} in 1991. It is thus natural to ask if similar results can be obtained when Newtonian gravity is replaced by general relativity. One should of course bear in mind that the situations are fundamentally different since in the latter case there exist data which lead to spacetime singularities, cf. \cite{AKR2,AR2,R2}. Nevertheless, as mentioned above, global existence may hold for polar or maximal time slicing.

The issue of global existence for the spherically symmetric Einstein-Vlasov system has previously been investigated in several papers, cf. \cite{RR,RRS,R1,A1,Ds2,AKR1,AKR2,AR1}. Global existence for small initial data has been proved in \cite{RR} and \cite{Ds2} for massive and massless particles respectively. In \cite{AKR2} initial data which guarantee formation of black holes are constructed, and it is proved that for a particular class of such initial data, with a steady state in the interior of the surrounding matter, global existence holds in Schwarzschild time. In \cite{AKR1} global existence is shown to hold in a maximal time gauge for rapidly outgoing matter. The methods of proofs in these cases are all tailored to treat special initial data and they will likely not apply in more general situations. The investigations \cite{RRS,R1,A1,AR1} are conditional in the sense that assumptions are made on the solutions, and not only on the initial data. However, the methods are general and cover large classes of initial data and can thus be thought of as possible approaches for treating the general case. The main purpose of the present work is to propose a method of approach for general initial data, which improves the regularity of the terms that need to be estimated compared to the methods in \cite{RRS,R1,A1,AR1}. Furthermore, the method improves and simplifies some of the previous results. To be more precise, let us discuss the relation between these studies. 

An important quantity in the study of the Einstein-Vlasov system and the Vlasov-Poisson system is
\begin{equation}\label{Qcompact}
Q(t):=\sup \{1+|v|:\exists (s,x)\in [0,t]\times \mathbb{R}^3\mbox{ such that
    }f(s,x,v)\not= 0\}.
\end{equation}
Here $f$ is the density function on phase, and $v\in\mathbb{R}^3$ is the momentum. $Q(t)$ measures the support of the momenta, and the content of the continuation criterion for these systems, cf. \cite{RR,R1,B}, is that solutions can be extended as long as $Q(t)$ remains bounded. The definition of $Q$ only applies in the case when the initial data have compact support in the momentum variables. For non-compact initial data, it was shown in \cite{AR1} that solutions can be extended as long as $\tilde{Q}(t)$ is bounded where

\begin{equation}\label{Qsupp}
\tilde{Q}(t)
:=\sup\left\{\frac{1+|V(s,0,r,v)|}{1+|v|} \mid 0\leq s\leq t,\
(r,v)\in \supp \fn \,\right\},
\end{equation}
Here $V$ is a solution of the characteristic system, cf. (\ref{char2}).
The analysis in \cite{RRS,A1,AR1} is carried out in Schwarzschild coordinates. The main result in \cite{RRS} shows that as long as there is no matter in the ball
$$\{x\in\mathbb{R}^3:|x|\leq\epsilon\},$$ the estimate
\begin{equation}\label{Qee}
Q(t)\leq e^{\log{Q(0)}e^{C(\epsilon)t}},
\end{equation}
holds. Here $C(\epsilon)$ is a constant which depends on $\epsilon.$ In view of the continuation criterion this can thus be viewed as a global existence result outside the centre of symmetry for initial data with compact support. The bound is obtained by estimating each term individually in the characteristic equation (\ref{char2}) for the radial momentum. This involves a particular difficulty. Let us consider the term involving $\mu_r$ in (\ref{char2}). The Einstein equations imply that
$$\mu_r=\frac{m}{r^2}e^{2\lambda}+4\pi rpe^{2\lambda}=:T_1+T_2,$$
where $m$ is the quasi local mass and $p$ is the pressure. There is a distinct difference between the terms $T_1$ and $T_2$ due to the fact that $m$ can be regarded as an average, since it is given as a space integral of the energy density $\rho,$ whereas $p$ is a pointwise term. Also the term involving $\lambda_t$ in (\ref{char2}) is a pointwise term in this sense. The method in \cite{RRS} is able to estimate the pointwise terms outside the centre but generally it seems very unpleasant to have to treat these terms. In the present work we give an alternative and simplified proof of the result in \cite{RRS}. In particular our method avoids the pointwise terms by using the fact that the characteristic system can be written in a form such that Green's formula in the plane can be applied. This results in a combination of terms involving second order derivatives which can be substituted for by one of the Einstein equations. This method was first introduced in \cite{A1} but here the set up is different and the application of Green's formula becomes very natural. In addition the bound of $Q$ is improved compared to (\ref{Qee}) and reads $$Q(t)\leq (Q(0)+\frac{Ct}{\epsilon^2})e^{C(1+t)/\epsilon}.$$ This bound implies that also $\tilde{Q}(t)$ is bounded and therefore a consequence of the method is that global existence outside the centre also holds for non-compact initial data. This improves the result in \cite{AR1} where only non-compact data for which the matter is ingoing, and such that the matter keeps on going inwards for all times, are admitted. Another consequence of our method is that global existence holds for ingoing matter as long as $3m(t,r)\leq r.$ Note that in Schwarzschild coordinates $2m(t,r)\leq r$ always, and that there are closed null geodesics if $3m=r$ in the Schwarzschild spacetime. This result was already proved in \cite{A1} but an additional assumption was imposed which now has been dropped.

We now turn to the case of maximal-isotropic coordinates. Rendall shows in \cite{R1} global existence outside the centre in maximal-isotropic coordinates. The bound on $Q(t)$ is again obtained by estimating each term in the characteristic equation, cf. equation (\ref{charW}). In this case there are no pointwise terms in contrast to the case with Schwarzschild coordinates. The terms are however, in analogy with the Schwarzschild case, strongly singular at the centre. The method that we use in Schwarzschild coordinates also applies in this case. The improvement lies in the fact that there is some gain in regularity, i.e. the terms that need to be estimated are less singular compared to the corresponding terms in \cite{R1}. Indeed, these terms can schematically, in both the Schwarzschild and maximal-isotropic case, be written as
\begin{equation}\label{g}
\int_0^{\infty}g(t,\eta)d\eta,
\end{equation}
and the known a priori bounds read
\begin{equation}\label{rg}
\int_0^r\eta g(t,\eta)d\eta.
\end{equation}
Hence, the degree of the singularity is of order one in the radial variable. Roughly, for the methods in \cite{RRS} and \cite{R1} the singularity is of second order. An advantage of our method is clearly that it applies in both cases, and it also turns out that the principal term to be estimated in the two cases is the spacetime integral of the Gauss curvature of the two dimensional quotient manifold ${\cal{M}}/SO(3),$ where ${\cal{M}}$ is the four dimensional spacetime manifold. This is interesting since the Gauss curvature is a coordinate independent quantity. Moreover, the principal term in \cite{Cu2}, p. 1172, has the same form. Finally we mention that our method also applies in the case of maximal-areal coordinates which were used in the study \cite{AKR1}. The analysis in this case is completely analogous and is left out.

The outline of the paper is as follows. The Einstein-Vlasov system in
Schwarzschild coordinates is treated in sections 2-5. In section 2 the system is formulated and
the a priori bounds are given in section 3. Section 4 is devoted to the proof of global existence outside the centre and in section 5 global existence is shown for ingoing matter which satisfies $m/r\leq 1/3.$ Sections 6-8 concern the system in maximal-isotropic coordinates. The system is given in section 6, and in section 7 the necessary a priori bounds are derived. Section 8 is devoted to the proof of the global existence theorem outside the centre.

\section{The Einstein-Vlasov system}
\setcounter{equation}{0}
For an introduction to the Einstein-Vlasov system and kinetic theory we refer to \cite{A2,R1}, and for a careful derivation of the system given below we refer to~\cite{R}.
In Schwarzschild coordinates the spherically symmetric metric takes the form
\begin{equation}
ds^{2}=-e^{2\mu(t,r)}dt^{2}+e^{2\lambda(t,r)}dr^{2}
+r^{2}(d\theta^{2}+\sin^{2}{\theta}d\varphi^{2}).
\end{equation}
The Einstein equations read
\begin{eqnarray}
&\displaystyle e^{-2\lambda}(2r\lambda_{r}-1)+1=8\pi r^2\rho,&\label{ee1}\\
&\displaystyle e^{-2\lambda}(2r\mu_{r}+1)-1=8\pi r^2 p,&\label{ee2}\\
&\displaystyle\lambda_{t}=-4\pi re^{\lambda+\mu}j,&\label{ee3}\\
&\displaystyle e^{-2\lambda}(\mu_{rr}+(\mu_{r}-\lambda_{r})(\mu_{r}+
\frac{1}{r}))-e^{-2\mu}(\lambda_{tt}+\lambda_{t}(\lambda_{t}-\mu_{t}))=
8\pi p_T.&\label{ee4}
\end{eqnarray}
The indicies $t$ and $r$ denote derivatives. 
The Vlasov equation for the density distribution function
$f=f(t,r,w,L)$ is given by
\begin{equation}
\partial_{t}f+e^{\mu-\lambda}\frac{w}{E}\partial_{r}f
-(\lambda_{t}w+e^{\mu-\lambda}\mu_{r}E-
e^{\mu-\lambda}\frac{L}{r^3E})\partial_{w}f=0,\label{vlasov}
\end{equation}
where
\begin{equation}
E=E(r,w,L)=\sqrt{1+w^{2}+L/r^{2}}.\label{E}
\end{equation}
Here $w\in (-\infty,\infty)$ can be thought of as the radial component
of the momentum variables, and $L\in [0,\infty)$ is the square of
the angular momentum.
The matter quantities are defined by
\begin{eqnarray}
\rho(t,r)&=&\frac{\pi}{r^{2}}
\int_{-\infty}^{\infty}\int_{0}^{\infty}Ef(t,r,w,L)\;dwdL,\label{rho}\\
p(t,r)&=&\frac{\pi}{r^{2}}\int_{-\infty}^{\infty}\int_{0}^{\infty}
\frac{w^{2}}{E}f(t,r,w,L)\;d
wdL,\label{p}\\
j(t,r)&=&\frac{\pi}{r^{2}}
\int_{-\infty}^{\infty}\int_{0}^{\infty}wf(t,r,w,L),\;dwdL,\label{j}\\
p_T(t,r)&=&\frac{\pi}{2r^{4}}\int_{-\infty}^{\infty}\int_{0}^{\infty}\frac{L}{E}f(t,r,w,L)\;
dwdL.\label{q}
\end{eqnarray}
Here $\rho,p,j$ and $p_T$ are the energy density, the radial pressure, the current and the tangential pressure respectively.
The following boundary conditions are imposed to ensure asymptotic flatness
\begin{equation}
\lim_{r\rightarrow\infty}\lambda(t,r)=\lim_{r\rightarrow\infty}\mu(t,r)=0,\label{bdryaf}
\end{equation}
and a regular centre requires
\begin{equation}\label{bdryrc}
\lambda(t,0)=0,\;t\geq 0.
\end{equation}
We point out that the Einstein equations are not independent and that
e.g. the equations (\ref{ee3}) and (\ref{ee4}) follow by
(\ref{ee1}), (\ref{ee2}) and (\ref{vlasov}).

As initial data it is sufficient to prescribe a distribution function
$\open{f}=\open{f}(r,w,L)\geq 0$ such that
\begin{equation}\label{notsinit}
   \int_0^r 4\pi\eta^2\open{\rho}\,(\eta)\,d\eta < \frac{r}{2}.
\end{equation}
Here we denote by $\open{\rho}$ the energy density induced by the initial
distribution function $\open{f}$. This condition ensures that no trapped surfaces are present
initially. Given $\open{f},$ equations (\ref{ee1}) and (\ref{ee2}) can be solved to give $\lambda$ and $\mu$ at $t=0.$ We will only consider initial data
such that $\fn=0$ if $L>L_2,$ for some $L_2>0,$
and such that $\fn=0$ if $r>R_2,$ for some $R_2>0.$
If in addition the initial data is $C^1$ we say that it is regular.
In most of the previous investigations the condition of compact support on the momentum
variable $w$ has
been included in the definition of regular data. The exception is \cite{AR1} where non-compact initial
data is studied and
a decay condition replaces the assumption of compact support. In this study we also
include data with non-compact support and we impose the decay condition from \cite{AR1}
\begin{equation}\label{iddecay}
\sup_{(r,w,L)\in\R^3} |w|^5 \fn (r,w,L) < \infty.
\end{equation}
We distinguish between the two cases; regular initial data with compact support
and regular initial data which satisfy the decay condition (\ref{iddecay}). We denote these
classes of initial data by $\cal{I}^C$ and $\cal{I}^D$ respectively.

The main results below concern subclasses of $\cal{I}^C$ and $\cal{I}^D.$ Given $R_{1}>0,$ we define
the subclass ${\cal{I}}^C(R_1)$ with radial cut-off by
\[
{\cal{I}}^C(R_1)=\{\fn\in{\cal{I}}^C: \fn=0\mbox{ for }r\leq R_{1}\}.
\]
The subclass ${\cal{I}}^D(R_1)$ is defined analogously.

Let us now write down a couple of
facts about the system (\ref{ee1})-(\ref{bdryrc}).
A solution to the Vlasov
equation can be written
\begin{equation}
f(t,r,w,L)=f_{0}(R(0,t,r,w,L),W(0,t,r,w,L),L),
\label{repre}
\end{equation}
where $R$ and $W$ are solutions of the characteristic system
\begin{eqnarray}
\frac{dR}{ds}&=&e^{(\mu-\lambda)(s,R)}\frac{W}{E(R,W,L)},\label{char1}\\
\frac{dW}{ds}&=&-\lambda_{t}(s,R)W-e^{(\mu-\lambda)(s,R)}\mu_{r}(s,R)E(R,W,L)
\nonumber\\
& &+e^{(\mu-\lambda)(s,R)}\frac{L}{R^3E(R,W,L)},\label{char2}
\end{eqnarray}
such that $(R(s,t,r,w,L),W(s,t,r,w,L),L)=(r,w,L)$ when $s=t$.
This representation shows that $f$ is nonnegative for all $t\geq 0,$ $\|f\|_{\infty}=\|f_{0}\|_{\infty},$ and that $f(t,r,w,L)=0$ if $L>L_{2}.$
The quasi local mass $m$ is defined by
\begin{equation}
m(t,r)=4\pi\int_{0}^{r}\eta^{2}\rho(t,\eta)d\eta,\label{m}
\end{equation}
and by integrating (\ref{ee1}) we find
\begin{equation}
e^{-2\lambda(t,r)}=1-\frac{2m(t,r)}{r}.\label{e2-lambda}
\end{equation}
A fact that we will need is that $$\mu+\lambda\leq 0.$$
This is easily seen by adding the equations (\ref{ee1}) and
(\ref{ee2}), which gives $$\lambda_{r}+\mu_{r}\geq 0,$$ and then using
the boundary conditions on $\lambda$ and $\mu.$ Furthermore, from (\ref{e2-lambda})
we get that $\lambda\geq 0,$ and it follows that $\mu\leq 0.$ We also introduce the notations
$\hat{\mu}$ and $\check{\mu}.$ From equation (\ref{ee2}) we have
\begin{equation}\label{muhatcheck}
\mu(t,r)=-\int_r^{\infty}\frac{m(t,\eta)}{\eta^2}e^{2\lambda}-\int_r^{\infty}4\pi \eta pe^{2\lambda}\, d\eta=:\hat{\mu}+\check{\mu}.
\end{equation}
Finally, we note that in~\cite{RR} and \cite{AR1} local existence theorems are proved for compact
and non-compact initial data respectively,
and it will be used below that solutions exist on some time interval $[0,T[.$

\section{A priori bounds in Schwarzschild coordinates}
\setcounter{equation}{0}
In this section we collect the a priori bounds that we need in the proofs below.
There are two known conserved quantities for the Einstein-Vlasov
system, the number of particles and the ADM mass $M$. Here, we will only need the latter
which is given by
\begin{equation}
M=4\pi\int_{0}^{\infty}r^{2}\rho(t,r)dr.\label{adm}
\end{equation}
The conservation of the ADM mass follows from general arguments but it can easily be
obtained by simply taking the time derivative of the integral expression and
use of the Vlasov equation and the Einstein equations.
The following results are given in \cite{A1} but since the proofs are very short we
have included them here for completeness. By a regular solution we mean a solution which
is launched by regular initial data.
\begin{lemma}
Let $(f,\mu,\lambda)$ be a regular solution to the Einstein-Vlasov system.
Then
\begin{eqnarray}
&\displaystyle\int_{0}^{\infty}4\pi r(\rho+p)e^{2\lambda}e^{\mu+\lambda}dr\leq 1,&\label{expr}\\
&\displaystyle\int_{0}^{\infty}(\frac{m}{r^2}+4\pi rp)e^{2\lambda}e^{\mu}dr\leq
1.&\label{expr2}
\end{eqnarray}
\end{lemma}
\textbf{Proof. }Using the boundary condition (\ref{bdryaf}) we get
\begin{eqnarray*}
1\geq
1-e^{\mu+\lambda}(t,0)&=&\int_{0}^{\infty}\frac{d}{dr}e^{\mu+\lambda}dr\\
&=&\int_{0}^{\infty}(\mu_{r}+\lambda_{r})e^{\mu+\lambda}dr.
\end{eqnarray*}
The right hand side equals (\ref{expr}) by equations (\ref{ee1})
and (\ref{ee2}) which completes the first part of the lemma. The
second part follows by studying $e^{\mu}$ instead of
$e^{\mu+\lambda}.$
\begin{flushright}
$\Box$
\end{flushright}
Next we show that not only $\rho(t,\cdot)\in L^{1},$ which follows from
the conservation of the ADM mass, but that also $e^{2\lambda}\rho(t,\cdot)\in
L^{1}.$
\begin{lemma}
Let $(f,\mu,\lambda)$ be a regular solution to the Einstein-Vlasov
system.
Then
\begin{equation}
\int_{0}^{\infty}r^{2}e^{2\lambda}\rho(t,r) dr\leq
\int_{0}^{\infty}r^{2}e^{2\lambda}\rho(0,r) dr+\frac{t}{8\pi}.\label{r2rhoe2lambda}
\end{equation}
\end{lemma}
\textbf{Proof. }Using the Vlasov equation we obtain
\begin{eqnarray*}
\partial_{t}\left(r^{2}e^{2\lambda}\rho(t,r)\right)&=&
-\partial_{r}\left(r^{2}e^{\mu+\lambda}j\right)-re^{\mu+\lambda}2je^{2\lambda}
\frac{m}{r}\\
&\leq&-\partial_{r}\left(r^{2}e^{\mu+\lambda}j\right)+
\frac{1}{2}re^{\mu+\lambda}(\rho+p)e^{2\lambda}.
\end{eqnarray*}
Here we used that $m/r\leq 1/2$ together with the elementary inequality
$2|j|\leq\rho+p,$ which follows from the expressions
(\ref{rho})-(\ref{j}). In view of (\ref{repre}) and (\ref{char1}) we see that
$\lim_{r\to\infty}r^2j(t,r)=0,$ since the initial data has compact
support in $r$. Since the solution is regular and hence bounded the boundary
term at $r=0$ also vanishes (as a matter of fact spherical symmetry
even implies that $j(t,0)=0$). Thus, by lemma 1 we get
\begin{displaymath}
\frac{d}{dt}\int_{0}^{\infty}r^{2}e^{2\lambda}\rho(t,r) dr\leq
\frac{1}{8\pi},
\end{displaymath}
which completes the proof of the lemma.
\begin{flushright}
$\Box$
\end{flushright}

\section{A regularity result in Schwarzschild coordinates}
\setcounter{equation}{0}
\begin{theorem}\label{theorem1} Let $0<\epsilon<R_1.$ Consider a solution of
the spherically symmetric
Einstein-Vlasov system, launched by initial data in ${\cal{I}}^D(R_1),$
on its maximal time interval $[0,T[$ of existence. If $f(s,r,\cdot,\cdot)=0,$ for
$(s,r)\in [0,t[\times [0,\epsilon],$ then
\begin{equation}\label{Qestimate}
\tilde{Q}(t)\leq (1+\frac{Ct}{\epsilon^2})e^{\frac{C(1+t)}{\epsilon}}.
\end{equation}
In particular, if $f(t,r,\cdot,\cdot)=0$ for $(t,r)\in [0,T[\times [0,\epsilon],$ then $T=\infty.$
\end{theorem}
The last statement in the theorem is a consequence of the continuation criterion derived in \cite{AR1}. The theorem thus improves the result in \cite{AR1} which was restricted to a special class of initial data where all the matter is ingoing. The result holds in particular for compactly supported data, and the bound (\ref{Qestimate}) improves the bound in \cite{RRS}, cf. inequality (\ref{Qee}).

\textit{Proof: }
We consider the quantities $G=E+W$ and $H=E-W,$ which satisfy $G>0,\; H>0.$
Along a characteristic $(R(s),W(s),L)$ we have by (\ref{char1}) and (\ref{char2}).
\begin{equation}
\frac{dG}{ds}=-\left[\lambda_t\frac{W}{E}+\mu_r e^{\mu-\lambda}\right]G
+\frac{Le^{\mu-\lambda}}{R^3E},\label{charG}
\end{equation}
and
\begin{equation}
\frac{dH}{ds}=\left[\lambda_t\frac{W}{E}+\mu_r e^{\mu-\lambda}\right]H
-\frac{Le^{\mu-\lambda}}{R^3E}.\label{charH}
\end{equation}
Let us first consider the quantity $H.$ From (\ref{charH}) we have
\begin{equation}
\frac{dH}{ds}\leq\left[\lambda_t\frac{W}{E}+\mu_r e^{\mu-\lambda}\right]H.\label{2charH}
\end{equation}
It follows that
\begin{equation}
H(t)\leq H(0)e^{\int_{0}^{t}\left[\lambda_t(s,R(s))\frac{W(s)}{E(s)}+\mu_r(s,R(s))e^{(\mu-\lambda)(s,R(s))}
\right]\, ds}.\label{HT}
\end{equation}
Let us denote the curve $(s,R(s)),\; 0\leq s\leq t,$ by $\gamma.$
By using that $$\frac{dR}{dt}=e^{\mu-\lambda}\frac{W}{E},$$
the integral above can be written as the curve integral
\begin{equation}
\int_{\gamma}e^{(-\mu+\lambda)(t,r)}\lambda_{t}(t,r)dr+
e^{(\mu-\lambda)(t,r)}\mu_{r}(t,r)dt.\label{curveH}
\end{equation}
Let $\Gamma$ denote the closed curve
$\Gamma:=\gamma+C_{t}+C_{\infty}+C_{0},$ oriented clockwise, where
\begin{eqnarray}
C_{t}&=&\{(t,r):R(t)\leq r\leq R_{\infty}\},\\
C_{\infty}&=&\{(s,R_{\infty}):t\geq s\geq 0\},\\
C_{0}&=&\{(0,r):R_{\infty}\geq r\geq R(0)\}.\label{Ccurves}
\end{eqnarray}
Here $R_{\infty}\geq R_{2}+t,$ so that $f=0$ when $r\geq R_{\infty}.$
We now apply Green's formula in the plane and use equation (\ref{ee4}) to obtain
\begin{eqnarray}
&\displaystyle\oint_{\Gamma} e^{-\mu+\lambda}\lambda_{t}dr+
e^{\mu-\lambda}\mu_{r}dt&\nonumber\\
&\displaystyle =\int\int_{\Omega}\partial_{t}\left(e^{-\mu+\lambda}
\lambda_{t}\right)-\partial_{r}
\left( e^{\mu-\lambda}\mu_{r}\right)dtdr&\nonumber\\
&\displaystyle=\int\int_{\Omega}\left(\lambda_{tt}+(\lambda_{t}-\mu_{t}\right)
\lambda_{t}-(\mu_{rr}+(\mu_{r}-\lambda_{r})\mu_{r})e^{\mu-\lambda}\,dtdr&
\nonumber\\
&\displaystyle =\int\int_{\Omega}(\mu_{r}-\lambda_{r})\frac{e^{\mu-\lambda}}{r}-8\pi
p_T e^{\mu+\lambda}\,dtdr&
\nonumber\\
&\displaystyle =\int\int_{\Omega}e^{\mu+\lambda}\left(\frac{2m}{r^{3}}
-4\pi(\rho-p)-8\pi p_T\right)\,dtdr.&\label{id}
\end{eqnarray}
Using the hypothesis that matter stays away from the region $r\leq\epsilon,$ we have $R(s)\geq\epsilon,\, 0\leq s\leq t,$ and we get in view of (\ref{expr2}), together with the facts that $\lambda\geq 0,\;\rho-p\geq 0$
and $q\geq 0$,
\begin{equation}\label{Omegaest}
\int\int_{\Omega}e^{\mu+\lambda}\left(\frac{2m}{r^{3}}
-4\pi(\rho-p)-8\pi p_T\right)\,dtdr\leq \frac{2t}{\epsilon}.
\end{equation}
\textbf{Remark 1: }The integrand above has a geometrical meaning. It is the scalar curvature of the quotient manifold ${\cal{M}}/SO(3)$ with metric $ds^2=-e^{2\mu}dt^2+e^{2\lambda}dr^2.$ It is interesting to note that the principal term in \cite{Cu2}, p. 1172, has the same form. \\
Now we wish to estimate the curve integral (\ref{curveH}). We have
\begin{equation}
\int_{\gamma}...=\oint_{\Gamma}...-\int_{C_t}4\pi rje^{2\lambda}dr-\int_{C_{\infty}}
\frac{M}{R_{\infty}^2}dt-\int_{C_0}4\pi rje^{2\lambda}dr.
\end{equation}
Here we used that $\lambda_t=4\pi rje^{\mu+\lambda}$
and that $\mu_r(t,R_{\infty})=\frac{M}{R_{\infty}^2}.$
The integral along $C_0$ is given in terms of the initial data,
\begin{equation}
\big|\int_{C_0}4\pi rje^{2\lambda}dr\big|\leq C.
\end{equation}
By letting $R_{\infty}\to\infty$ the integral along $C_{\infty}$ vanishes and it remains to
estimate the contribution from the integral along $C_t.$ In view of (\ref{r2rhoe2lambda})
we get
\begin{equation}\label{intrj}
\big|\int_{C_t}4\pi rje^{2\lambda}dr\big|\leq \frac{1}{\epsilon}\int_0^{\infty}
4\pi r^2\rho(t,r)e^{2\lambda(t,r)}dr\leq \frac{C}{\epsilon}(1+t).
\end{equation}
The esimate of $G$ is very similar. Indeed, by the hypothesis of the theorem we
have $R(t)\geq \epsilon,$ thus
\begin{equation}
\frac{Le^{\mu-\lambda}}{R^3E}\leq \frac{\sqrt{L}}{\epsilon^2}
\leq\frac{\sqrt{L_2}}{\epsilon^2}.
\end{equation}
We derive in view of (\ref{charG})
\begin{eqnarray}
\displaystyle G(t)&\leq& G(0)e^{\int_{0}^{t}\left[-\lambda_t(s,R(s))
\frac{W(s)}{E(s)}-\mu_r(s,R(s))e^{(\mu-\lambda)(s,R(s))}\right]ds}\nonumber\\
&+&\int_0^t\frac{\sqrt{L_2}}{\epsilon^2}e^{\int_{\tau}^{t}\left[-\lambda_t(s,R(s))\frac{W(s)}{E(s)}-\mu_r(s,R(s))e^{(\mu-\lambda)(s,R(s))}\right]ds}d\tau.\nonumber
\end{eqnarray}
Note that the integrals in the exponent above are identical to the integral in (\ref{HT}) except for the sign.
We can accordingly use the same arguments as above, the only difference is that we use 
(\ref{intrj}) twice, for $C_t$ and $C_{\tau}$, and that the integral in (\ref{id}) has 
opposite sign. Since $2p_T\leq \rho-p$ it is sufficient to estimate 
the integral
\begin{equation}
\int\int_{\Omega} 8\pi(\rho-p)e^{\mu+\lambda}\,dtdr.
\end{equation}
We use the bound (\ref{expr}) to obtain the estimate (\ref{Omegaest}) also in this case. 
Hence, we have the following bounds
\begin{equation}
G(t)\leq (G(0)+\frac{Ct}{\epsilon^2})e^{\frac{C(1+t)}{\epsilon}},
\end{equation}
and
\begin{equation}
H(t)\leq H(0)e^{\frac{C(1+t)}{\epsilon}}.
\end{equation}
Since $E=G+H$ this implies that
\begin{equation}
\tilde{Q}(t)\leq (1+\frac{Ct}{\epsilon^2})e^{\frac{C(1+t)}{\epsilon}}.
\end{equation}
This completes the proof of Theorem \ref{theorem1}.
\begin{flushright}
$\Box$
\end{flushright}

\section{Global existence for ingoing matter with $3m\leq r$}
\setcounter{equation}{0}
In this section we consider compactly supported initial data $\open{f}\in{\cal{I}}^C(R_1)$ such that for given $0<L_1<L_2$ and $P>0,$ it holds that $L\geq L_1,$ and  $w<-P$ for all $(w,L)\in\supp\open{f}.$ Furthermore, the initial data have the property that $3m<r$ everywhere. Note that in Schwarzschild spacetime there are closed null geodesics when $3m=r.$ Let us denote this class of initial data by ${\cal{I}}^C(R_1,L_1,P,3)\subset{\cal{I}}^C(R_1).$ By continuity there is a $T_1>0$ such that $w\leq 0$ for $w\in\supp f(t),$ and $3m(t,r)\leq r$ everywhere, for $t\leq T_1.$ We will show that on the time interval $[0,T_1]$ singularities do not form in the evolution. This can be phrased as a global existence result for ingoing matter satisfying $3m\leq r$. A similar result is proved in
\cite{A1} but an additional assumption on the solution is imposed and the present result is thus an improvement.
\begin{theorem}\label{theorem2}
Consider a solution to the spherically symmetric Einstein-Vlasov system launched by initial data
$\open{f}\in {\cal{I}}^C(R_1,L_1,P,3)$. Let $T>0$ be the maximal time interval on
which the solution exists and let $T_1$ be as above. If $T<\infty$ then $T_1<T.$
\end{theorem}
\textit{Proof: }
Let us consider the quantity $He^{\hat{\mu}}$ along a given characteristic $(t,R(t),W(t),L),$
with $W\leq 0.$
In view of (\ref{charH}) and (\ref{muhatcheck}) we have
\begin{equation}
\frac{d}{ds}(He^{\hat{\mu}})=\left[\lambda_t\frac{W}{E}+\mu_r e^{\mu-\lambda}+\hat{\mu}_t+\hat{\mu_r}\frac{W}{E}e^{\mu-\lambda}\right]He^{\hat{\mu}}
-\frac{Le^{\mu-\lambda}}{R^3E}e^{\hat{\mu}}.\label{charHexpmuhat}
\end{equation}
Since $\mu=\hat{\mu}+\check{\mu}$ we have
\begin{equation}
\mu_r e^{\mu-\lambda}+\hat{\mu_r}\frac{W}{E}e^{\mu-\lambda}=\check{\mu}_r e^{\mu-\lambda}
+\hat{\mu}_r e^{\mu-\lambda}(1+\frac{W}{E})=\check{\mu}_r e^{\mu-\lambda}
+\hat{\mu}_r e^{\mu-\lambda}\frac{1+L/R^2}{E(E-W)}.
\end{equation}
Since $H=E-W$ and $$\hat{\mu}_r=\frac{me^{2\lambda}}{r^2},$$ we obtain
\begin{equation}
\frac{d}{ds}(He^{\hat{\mu}})=\left[\lambda_t\frac{W}{E}+\check{\mu}_r e^{\mu-\lambda}+\hat{\mu}_t
\right]He^{\hat{\mu}}
+e^{\mu-\lambda}
\left(\frac{me^{2\lambda}}{R}\frac{(R^2+L)}{R^3E}-\frac{L}{R^3E}\right)e^{\hat{\mu}}.
\label{charHexpmuhatweak}
\end{equation}
We note that if
\begin{equation}\label{weak}
\frac{m(t,R(t))}{R(t)}\leq\frac{1}{3},
\end{equation}
we have
\begin{equation}
\frac{m(t,R(t))}{R(t)}e^{2\lambda(t,R(t))}=\frac{m(t,R(t))}{R(t)(1-\frac{2m(t,R(t))}{R(t)})}\leq 1.
\end{equation}
Hence, as long as (\ref{weak}) holds true we get in view of (\ref{charHexpmuhatweak})
the inequality
\begin{equation}
\frac{d}{ds}(He^{\hat{\mu}})\leq
\left[\lambda_t\frac{W}{E}+\check{\mu}_r e^{\mu-\lambda}+\hat{\mu}_t
\right]He^{\hat{\mu}}
+e^{\mu-\lambda}
\frac{1}{RE}e^{\hat{\mu}}.
\label{charHexpmuhatweakineq}
\end{equation}
Since $ER\geq\sqrt{L_1}$ the last term is bounded. We can without loss of generality 
assume that $He^{\hat{\mu}}\geq 1$ on $[0,T_1]$, and we get for $t\leq T_1,$
\begin{eqnarray}
\displaystyle H(t)e^{\hat{\mu}(t,R(t))}&\leq &H(0)e^{\hat{\mu}(0,R(0))}
e^{\frac{t}{\sqrt{L_1}}+\int_0^t \hat{\mu}_t(s,R(s))ds}\nonumber\\
\displaystyle & &\times e^{\int_{0}^{t}
\left[\lambda_t(s,R(s))\frac{W}{E}
+\check{\mu}_r(s,R(s))e^{(\mu-\lambda)(s,R(s))}\right]ds}.\label{Hexphatmuweak}
\end{eqnarray}
Let us denote by $\gamma$ the curve $(t,R(t)),\, 0\leq t\leq T_1.$
The last integral in (\ref{Hexphatmuweak}) can be written as
\begin{equation}
\int_{\gamma}e^{(-\mu+\lambda)(t,r)}\lambda_{t}(t,r)dr+
e^{(\mu-\lambda)(t,r)}\check{\mu}_{r}(t,r)dt.\label{Hcurveweak}
\end{equation}
We will apply the Green formula to this curve integral and we introduce as above
the closed curve $\Gamma=\gamma+C_{T_1}+C_{\infty}+C_0$ where $C_{T_1},\,C_{\infty}$ and
$C_0$ are defined as in (\ref{Ccurves}).
We have
\begin{eqnarray}
&\displaystyle\oint_{\Gamma} e^{-\mu+\lambda}\lambda_{t}dr+
e^{\mu-\lambda}\check{\mu}_{r}ds&\nonumber\\
&\displaystyle =\int\int_{\Omega}\partial_{t}\left(e^{-\mu+\lambda}
\lambda_{t}\right)-\partial_{r}
\left( e^{\mu-\lambda}\check{\mu}_{r}\right)dsdr&\nonumber\\
&\displaystyle =\int\int_{\Omega}\partial_{t}\left(e^{-\mu+\lambda}
\lambda_{t}\right)-\partial_{r}
\left( e^{\mu-\lambda}\mu_{r}\right)dsdr
+\int\int_{\Omega}\partial_{r}
\left(e^{\mu+\lambda}\frac{m}{r^{2}}\right)dsdr.&\nonumber\\
\end{eqnarray}

For the first integral above we use (\ref{id}) and we obtain
the identity
\begin{eqnarray}
& &\phantom{GH} \displaystyle\oint_{\Gamma} e^{-\mu+\lambda}\lambda_{t}dr+
e^{\mu-\lambda}\check{\mu}_{r}ds\nonumber\\
& &\displaystyle =\int\int_{\Omega}e^{\mu+\lambda}\left(\frac{2m}{r^{3}}
-4\pi(\rho-p)-8\pi p_T\right)\,dtdr\nonumber\\
& &\phantom{G} \displaystyle +\int\int_{\Omega}e^{\mu+\lambda}\left((\mu_r+\lambda_r)\frac{m}{r^2}+4\pi\rho-\frac{2m}{r^3}\right)dtdr\nonumber\\
& &\displaystyle=\int\int_{\Omega}4\pi e^{\mu+\lambda}\left[
(\rho+p)e^{2\lambda}\frac{m}{r}+2p-\rho\right] drds\nonumber\\
& &\phantom{H} \displaystyle+\int\int_{\Omega}\int_{-\infty}^{\infty}\int_{0}^{\infty}
\frac{4\pi^{2} e^{\mu+\lambda}}{r^{2}E}f(t,r,w,F)dFdw\,drds.\label{id2}
\end{eqnarray}
Here we used that $\mu_r+\lambda_r=4\pi r(\rho+p)e^{2\lambda}$ and that
\begin{equation}\label{almostq}
2p_T=\rho-p-\int_{-\infty}^{\infty}\int_{0}^{\infty}
\frac{4\pi^{2} e^{\mu+\lambda}}{r^{2}E}f(t,r,w,F)dFdw.
\end{equation}
Recall from (\ref{Hexphatmuweak}) that the integral involving $\hat{\mu}$ should also
be taken into account. Since
\begin{equation}
\hat{\mu}_{t}(t,r)=\int_r^{\infty}4\pi je^{2\lambda}e^{\mu+\lambda}d\eta,
\end{equation}
we get
\begin{equation}
\int_0^t\hat{\mu}_{t}(s,R(s))ds=\int\int_{\Omega}4\pi je^{2\lambda}e^{\mu+\lambda}dsdr.
\end{equation}
Hence we obtain with $C_B:=C_{T_1}+C_{\infty}+C_0,$
\begin{eqnarray}
&\displaystyle\oint_{\gamma} e^{-\mu+\lambda}\lambda_{t}dr+
e^{\mu-\lambda}\check{\mu}_{r}ds+\int_{0}^{T}\hat{\mu}_t(s,r) ds&\nonumber\\
&\displaystyle=\int\int_{\Omega}4\pi e^{\mu+\lambda}\left[
(\rho+p)e^{2\lambda}\frac{m}{r}+2p-\rho+je^{2\lambda}\right] drds&\nonumber\\
&\displaystyle+\int\int_{\Omega}\int_{-\infty}^{\infty}\int_{0}^{\infty}
\frac{4\pi^{2} e^{\mu+\lambda}}{r^{2}E}f(t,r,w,F)dFdw\,drds&\nonumber\\
&\displaystyle -\int_{C_B}e^{-\mu+\lambda}\lambda_{t}dr+
e^{\mu-\lambda}\check{\mu}_{r}ds.&\label{green3}
\end{eqnarray}
The integral over $C_B$ equals
\begin{eqnarray}
&\displaystyle-\int_{C_B}e^{-\mu+\lambda}\lambda_{t}dr+
e^{\mu-\lambda}\check{\mu}_{r}ds=\int_{R(T_1)}^{\infty} 4\pi rj(T_1,r)e^{2\lambda(T_1,r)}dr&
\nonumber\\
&\displaystyle+\int_{0}^{T_1}4\pi R_{\infty}p(s,R_{\infty})e^{(\mu+\lambda)(s,R_{\infty})}ds-
\int_{R(0)}^{\infty}4\pi rj(0,r)e^{2\lambda(0,r)}dr.&
\end{eqnarray}
The first integral on the right hand side is negative since matter is ingoing,
the second vanishes when $R_{\infty}$ is large since $p$ has compact support
and the third integral depends only on the initial data.
For the second integral in (\ref{green3}) we note that
\begin{equation}
\int_{-\infty}^{\infty}\int_{0}^{\infty}
\frac{4\pi^2 e^{\mu+\lambda}}{r^{2}E}f(t,r,w,F)dFdw\leq\frac{4\pi}{L_{1}}r^2\rho,
\end{equation}
which in view of (\ref{adm}) implies that it is bounded. It remains to show that the first
integral in (\ref{green3}) is bounded. 
Since $1+2\frac{m}{r}e^{2\lambda}=e^{2\lambda}$ we find that the integrand is non-positive
\begin{eqnarray*}
&\displaystyle(\rho+p)e^{2\lambda}\frac{m}{r}+2p-\rho+
je^{2\lambda}&\nonumber\\
&\displaystyle =
-(\rho-p)(1-e^{2\lambda}\frac{m}{r})+(p+j)e^{2\lambda}\leq
0.&\nonumber
\end{eqnarray*}
Here we used that $p\leq\rho, \;e^{2\lambda}m/r\leq 1,$ and that
$p+j\leq 0$ in view of (\ref{p}) and (\ref{j}) since matter is ingoing.
Hence $He^{\hat{\mu}}$ is bounded on $[0,T_1].$ Now, from
the characteristic equation (\ref{char1}) we have that
\begin{equation}
\frac{d}{ds}R^{-1}=\frac{-W}{R^2E}e^{\mu-\lambda}\leq \frac{-W}{R\sqrt{L_1}}e^{\hat{\mu}}\leq
\frac{He^{\hat{\mu}}}{R\sqrt{L_1}}\leq C(T_1)R^{-1}.
\end{equation}
It follows that for any characteristic $(t,R(t),W(t)),\;R(T_1)\geq\epsilon,$ for some $\epsilon>0,$ and Theorem \ref{theorem1}
thus applies which shows that the solution can be extended beyond $t=T_1.$ Hence, if $T<\infty,$ then  $T_1<T,$ and the proof of Theorem \ref{theorem2} is complete.
\begin{flushright}
$\Box$
\end{flushright}

\section{The Einstein-Vlasov system in maximal-isotropic coordinates}
\setcounter{equation}{0}
In \cite{R1} the Einstein-Vlasov system is studied in maximal-isotropic coordinates where the metric reads
\[
ds^2=-\alpha(t,R)dt^2+A^2(t,R)[(dR+\beta(t,R)dt)^2+R^2(d\theta^2+\sin^2{\theta}d\phi^2)].
\]
The condition that the hypersurfaces of constant time are maximal implies, cf. \cite{R1}, that the field equations take the following form:

\begin{eqnarray}
& &\partial_R(R^2\frac{A_{R}}{2\sqrt{A}})=
-\frac18A^{5/2}R^2(\frac32K^2+16\pi\rho)\label{A}\\
& &\alpha_{RR}+\frac{2}{R}\alpha_{R}+\frac{1}{A}A_{R}\,\alpha_{R}=\alpha A^2(\frac32 K^2+4\pi(\rho+p+2p_T))\label{alpha}\\
& &K_{R}+3(\frac{1}{A}A_{R}+\frac{1}{R})K=8\pi j\label{K}\\
& &\beta_{R}-\frac{1}{R}\beta=\frac32 \alpha K\label{beta}\\
& &A_t=-\alpha KA+\partial_R(\beta A)\label{At}\\
& &K_t=-\frac{1}{A^2}\alpha_{RR}+\frac{1}{A^3}A_{R}\alpha_{R}
+\beta K_{R}+4\pi\alpha(2p_T-p-\rho)\nonumber\\
& &\phantom{555555}
+\alpha[-\frac{2}{A^3}A_{RR}+\frac{2}{A^4}(A_{R})^2-\frac{2}{RA^3}A_{R}]\label{Kt}
\label{Kt}
\end{eqnarray}
The indicies $t$ and $R$ denote derivatives but sometimes we also use $\partial_t$ and $\partial_R$. 
In \cite{R1} the Vlasov equation is written in different coordinates than we use here. The relation is as follows. Let $x^{i}$ be the coordinates $(R\sin{\theta}\cos{\phi},R\sin{\theta}\cos{\phi},R\cos{\theta})$ and define an orthonormal frame by $e_i=A^{-1}\partial/\partial x^{i}.$ In \cite{R1} the mass shell is coordinatized by $(t,x^{a},v^{i}),$ where $v^{i}$ denote the components of a vector in the orthonormal frame, and the Vlasov equation in \cite{R1} is accordingly written in these variables. Now, let $w=(v\cdot x)/R,$ and let $L=A^2R^2(v^2-(x\cdot v)^2/R^2),$ then the Vlasov equation for the density distribution function $f=f(t,r,w,L)$ takes the form
\begin{equation}
\partial_{t}f+(\frac{\alpha}{A}\frac{w}{E}-\beta)\partial_{R}f
+(-\frac{E\alpha_{R}}{A}+\alpha Kw+\frac{\alpha L}{EA^3R^2}(\frac{1}{R}+\frac{A_{R}}{A}))
\partial_{w}f=0,\label{vlasovmi}
\end{equation}
where
\begin{equation}
E=E(R,w,L)=\sqrt{1+w^{2}+\frac{L}{A^2R^{2}}}.\label{E}
\end{equation}
The matter quantities are defined by
\begin{eqnarray}
\rho(t,R)&=&\frac{\pi}{A^2R^{2}}
\int_{-\infty}^{\infty}\int_{0}^{\infty}Ef(t,r,w,L)\;dwdL,\label{rhomi}\\
p(t,R)&=&\frac{\pi}{A^2R^{2}}\int_{-\infty}^{\infty}\int_{0}^{\infty}
\frac{w^{2}}{E}f(t,r,w,L)\;d
wdL,\label{pmi}\\
j(t,R)&=&\frac{\pi}{A^2R^{2}}
\int_{-\infty}^{\infty}\int_{0}^{\infty}wf(t,r,w,L),\;dwdL,\label{jmi}\\
p_T(t,R)&=&\frac{\pi}{2A^4R^{4}}\int_{-\infty}^{\infty}\int_{0}^{\infty}\frac{L}{E}f(t,r,w,L)\;
dwdL.\label{qmi}
\end{eqnarray}
To ensure asymptotical flatness we impose the boundary conditions
\begin{equation}\label{bdrys}
\lim_{R\to\infty}A(t,R)=1,\;\lim_{R\to\infty}\alpha(t,R)=1,\;\lim_{R\to\infty}\beta(t,R)=0,
\end{equation}
and to ensure a regular centre we require that
\begin{equation}\label{regularctr}
A(t,0)=1,\;\;\forall t\geq 0.
\end{equation}
In the case of Schwarzschild coordinates the constraint equations are easily solved for any given $\open{f}\in{\cal{I}}^D.$ This is not the case in maximal-isotropic coordinates. Here we assume that initial data are given, such that the constraint equations are satisfied, and such that $0\leq\open{f}\in C^1$ has compact support, $\open{f(R,w,L)}=0$ whenever $R<R_1, \;R>R_2$ or if $L>L_2,$ for given $0<R_1<R_2<\infty,$ and $L_2>0.$ Hence, we assume  properties which are similar to the initial data class ${\cal{I}}^C(R_1),$ with the notable difference that the condition (\ref{notsinit}) is not imposed since in these coordinates trapped surfaces are admitted. Let us denote this class of initial data by ${\cal{J}}^C(R_1).$

For a given metric the solution of the Vlasov equation is given by
\begin{equation}
f(t,R,w,L)=f_0({\cal{R}}(0,t,R,w,L),{\cal{W}}(0,t,R,w,L)),
\end{equation}
where $\cal{R}$ and $\cal{W}$ are solutions of the characteristic system
\begin{eqnarray}
\frac{d\cal{R}}{ds}&=&(\frac{\alpha}{A}\frac{\cal{W}}{E}-\beta),\label{charGamma}\\
\frac{d\cal{W}}{ds}&=&-\frac{E\alpha_{R}}{A}+\alpha K{\cal{W}}
+\frac{\alpha L}{EA^{3}{\cal{R}}^2}(\frac{1}{\cal{R}}+\frac{A_{R}}{A}).\label{charW}
\end{eqnarray}
Here ${\cal{R}}(t,t,R,w,L)=R,$ and ${\cal{W}}(t,t,R,w,L)=w.$

\section{A priori bounds in maximal-isotropic coordinates}
\setcounter{equation}{0}
In this section we collect the bounds needed in the next section.
The following estimates are derived in \cite{R1} by using the results of \cite{MO},
\begin{equation}\label{os}
A|K|\leq\frac{2}{R},\;\;\frac{|A_{R}|}{A}\leq\frac{2}{R}.
\end{equation}
Moreover, it follows from equation (\ref{beta}), cf. \cite{R1}, that
\begin{equation}\label{betaestimate}
|\beta|\leq 3.
\end{equation}
Another useful bound follows from conservation of the ADM mass $M,$ which is given by, cf. \cite{R1},
\begin{equation}\label{Mint}
M=\frac14\int_0^{\infty}A^{5/2}R^2(\frac32 K^2+16\pi\rho)\,dR.
\end{equation}
The following estimate plays the role of Lemma 1 in Schwarzschild coordinates.
\begin{lemma}\label{lemma4mi}
Let $(f,\alpha,A,\beta)$ be a regular solution of the Einstein-Vlasov system (\ref{A})-(\ref{regularctr}). Then
\begin{equation}
\int_0^{\infty}\alpha A^2\eta(\frac32 K^2+4\pi(\rho+p+2p_T))\,d\eta\leq 1.
\end{equation}
\end{lemma}
\textit{Proof: }Consider the second order equation (\ref{alpha}) for $\alpha$. The left hand side of
(\ref{alpha}) equals
\[
\frac{1}{R^2A}\partial_R(R^2A\alpha_{R}),
\]
which implies that
\[
\alpha_{R}(R)=\frac{1}{R^2A}\int_0^R\alpha A^3\eta^2(\frac32 K^2+4\pi(\rho+p+2p_T))\,d\eta.
\]
By using the boundary condition $\alpha(t,\infty)=1$ we get
\begin{equation}
1\geq 1-\alpha(0)=\int_0^{\infty}\frac{1}{R^2A}\big(\int_0^R \alpha A^3\eta^2(\frac32 K^2+4\pi(\rho+p+2p_T))\,d\eta\big)dR.
\end{equation}
From equation (\ref{A}) we get that $A$ is monotonic and decreasing. Changing the order of integration
and using the monotonicity of $A$ we get
\begin{eqnarray}\label{alphaA}
1&\geq&\int_0^{\infty}\big(\int_{\eta}^{\infty}\frac{1}{R^2A(\eta)}dR\big)\alpha A^3\eta^2(\frac32 K^2+4\pi(\rho+p+2p_T))\,d\eta\nonumber\\
&=&\int_0^{\infty}\alpha A^2\eta(\frac32 K^2+4\pi(\rho+p+2p_T))\,d\eta.
\end{eqnarray}
\begin{flushright}
$\Box$
\end{flushright}
We also need to establish a decay result for large $R.$
\begin{lemma}\label{lemmadecay}
A solution of the Einstein-Vlasov system (\ref{A})-(\ref{regularctr}) satisfies
\begin{equation}
\int_0^t\frac{\alpha_R(s,R)}{A(s,R)}\,ds\to 0\mbox{ as }R\to\infty.
\end{equation}
\end{lemma}
\textit{Proof: }
Using (\ref{Mint}) together with equation (\ref{A}) we obtain
\[
\frac{A_R}{\sqrt{A}}\geq -\frac{M}{R^2},
\]
which in view of the boundary condition $A(\infty)=1$ implies that
\begin{equation}\label{Aestimate}
A(t,R)\leq (1+\frac{M}{2R})^2.
\end{equation}
From (\ref{charGamma}) and (\ref{betaestimate}) we obtain that all characteristics originating
from the support of the matter have $R(t)\leq R_2+4t=:R_3(t).$
From equation (\ref{K}) we have in view of (\ref{os})
\begin{eqnarray}\label{RAK}
|(R^3A^3K)(t,R)|&=&\big|\int_0^R 8\pi\eta^3 A^4 j\,d\eta\big|
=\big|\int_0^{R_3(t)} 8\pi\eta^3 A^4 j\,d\eta\big|\nonumber\\
&=&|(R_{3}^3A^3K)(t,R_3(t))|\leq 2R_3(t)^2A^2(t,R_3(t)).
\end{eqnarray}
From (\ref{Aestimate}) we have
\[
A(t,R_3(t))\leq (1+\frac{M}{2R_2})^2\leq C.
\]
Thus, since $A\geq 1,$
\begin{equation}\label{Kestimate}
K(R)\leq \frac{CR_3(t)^2}{R^3}=C\frac{(R_2+4t)^2}{R^3}.
\end{equation}
Now we use equations (\ref{Kt}) and (\ref{At}) to derive
\begin{equation}
\partial_t(AKR)=-\frac74 RK^2A+RKA\beta+\frac{2\alpha_R}{A}(1+\frac{A_R}{A}+\frac{A_R^2}{2A^2})+\frac{2\alpha A_R}{A^2},
\end{equation}
where we used (\ref{A}) and (\ref{alpha}) to substitute the second order derivatives of $A$ and $\alpha.$
Since
\[
|\frac{A_R}{\sqrt{A}}|\leq\frac{M}{R^2},
\]
we have for sufficiently large $R$ that
\[
0\geq \frac{A_R}{A}+\frac{A_R^2}{2A^2}\geq -\frac12.
\]
Thus we get
\begin{equation}
0\leq\frac{\alpha_R}{A}\leq \partial_t(AKR)+\frac74 RK^2A-RKA\beta-\frac{2\alpha A_R}{A^2}.
\end{equation}
In view of the decay estimates for $A,\,A_{R}$ and $K$ and the bound of $\beta,$ we obtain
\[
\int_0^t\frac{\alpha_R(s,R)}{A(s,R)}\,ds\to 0\mbox{ as }R\to \infty.
\]
This completes the proof of the lemma.
\begin{flushright}
$\Box$
\end{flushright}

\section{A regularity result in maximal-isotropic coordinates}
\setcounter{equation}{0}
\begin{theorem}\label{theorem3} Let $0<\epsilon<R_1.$ Consider a solution of
the spherically symmetric
Einstein-Vlasov system, launched by initial data in ${\cal{J}}^C(R_1),$
on its maximal time interval $[0,T[$ of existence. If $f(s,R,\cdot,\cdot)=0,$ for
$(s,R)\in [0,t[\times [0,\epsilon],$ then
\begin{equation}\label{Qestimatemi}
Q(t)\leq (Q(0)+\frac{Ct}{\epsilon^2})e^{\frac{C(1+t^2)}{\epsilon}}.
\end{equation}
In particular, if $f(t,R,\cdot,\cdot)=0$ for $(t,R)\in [0,T[\times [0,\epsilon],$ then $T=\infty.$
\end{theorem}
The last statement in the theorem is a consequence of the continuation criterion
derived in \cite{R1}. Indeed, the requirements are that $Q(t)$ and $A(t,0)$ are bounded. Now, since there is no matter in the domain $R\leq\epsilon,$ a bound on $A(t,0)$ follows from equation (\ref{A}) and (\ref{Mint}), cf. the bound (\ref{Aestimate}). The result of the theorem is not new, it is included in \cite{R1}, but as was mentioned in the introduction the method is different and the terms that need to be estimated, in order to obtain global existence without assuming a lower bound of $R,$ are more regular in our approach.\\

\textit{Proof: }The method of proof is to a large extent analogous to the proof in Schwarzschild coordinates. Consider the quantities $G=E(t,{\cal{R}},{\cal{W}})+{\cal{W}}>0$ and
$H=E(t,{\cal{R}},{\cal{W}})-{\cal{W}}>0.$
Along a characteristic $({\cal{R}}(t),{\cal{W}}(t),L)$ we have by
(\ref{charGamma}) and (\ref{charW})
\begin{equation}\label{Gmi}
\frac{dG}{ds}=-[\frac{\alpha_{R}}{A}-\frac{\alpha K{\cal{W}}}{E}]G-\frac{L\alpha K}{2EA^2{\cal{R}}^2}
+\frac{\alpha L}{A^3{\cal{R}}^2E}(\frac{1}{{\cal{R}}}+\frac{A_{R}}{A}),
\end{equation}
and
\begin{equation}\label{Hmi}
\frac{dH}{ds}=[\frac{\alpha_{R}}{A}-\frac{\alpha K{\cal{W}}}{E}]H-\frac{L\alpha K}{2EA^2{\cal{R}}^2}
-\frac{\alpha L}{A^3{\cal{R}}^2E}(\frac{1}{{\cal{R}}}+\frac{A_{R}}{A}).
\end{equation}
First we consider the quantity $H.$
Using equation (\ref{os}) we conclude that
\[
\big|-\frac{L\alpha K}{2EA^3{\cal{R}}^2}
-\frac{\alpha L}{A^3{\cal{R}}^2E}(\frac{1}{{\cal{R}}}+\frac{A_{R}}{A})\big|\leq \frac{C\sqrt{L}}{{\cal{R}}^2}\leq \frac{C}{\epsilon^2}.
\]
Thus

\begin{equation}\label{Hest}
H(t)\leq H(0)e^{\int_0^t \frac{\alpha_{R}}{A}-\frac{\alpha K{\cal{W}}}{E} ds}+ \int_0^t\frac{C}{\epsilon^2}e^{\int_{\tau}^t \frac{\alpha_{R}}{A}-\frac{\alpha K{\cal{W}}}{E} ds}d\tau.
\end{equation}
Let us consider the first of the integrals in the exponent, the second is analogous.
Since
\[
\frac{d{\cal{R}}}{ds}=(\frac{\alpha}{A}\frac{{\cal{W}}}{E}-\beta),
\]
we can write the integral as a curve integral
\[
\int_{\gamma}-(KA)(t,R)dR+(\frac{\alpha_{R}(t,R)}{A(t,R)}-(\beta KA)(t,R))dt,
\]
where $\gamma$ is the curve $(s,{\cal{R}}(s)),\; 0\leq s\leq t.$
Let $\Gamma$ denote the closed curve
$\Gamma:=\gamma+C_{t}+C_{\infty}+C_{0},$ oriented clockwise, where
\begin{eqnarray}
C_{t}&=&\{(t,r):R(t)\leq r\leq R_{\infty}\},\\
C_{\infty}&=&\{(s,R_{\infty}):t\geq s\geq 0\},\\
C_{0}&=&\{(0,r):R_{\infty}\geq r\geq R(0)\}.\label{Ccurves}
\end{eqnarray}
Here $R_{\infty}\geq R_{2}+4t,$ so that $f=0$ when $r\geq R_{\infty}.$
By applying Green's formula in the plane we get
\begin{eqnarray}
&\displaystyle\oint_{\Gamma}-(KA)(t,R)dR
+(\frac{\alpha_{R}(t,R)}{A(t,R)}-(\beta KA)(t,R))dt&\nonumber\\
&\displaystyle =\int\int_{\Omega}\partial_{t}\left(-(KA)(t,R)\right)-\partial_{R}
\left(\frac{\alpha_{R}(t,R)}{A(t,R)}-(\beta KA)(t,R)\right)dtdR=:I_{\Omega}.&\nonumber
\end{eqnarray}
Now we use equation (\ref{Kt}) to substitute for $\partial_tK$ above. We obtain
\begin{equation}
\displaystyle I_{\Omega}=\int\int_{\Omega} \alpha A\big(\frac{2A_{RR}}{A^3}
-\frac{2(A_{R})^2}{A^4}+\frac{2A_{R}}{RA^3}+4\pi(\rho+p-2p_T)\big)+\alpha K^2A
\,dRdt.
\end{equation}
From (\ref{A}) we derive
\[
\frac{A_{RR}}{A^3}=\frac{(A_R)^2}{A^4}
-\frac{A_R}{RA^3}-\frac34 K^2-8\pi\rho.
\]
We thus get
\begin{equation}\label{Iomega}
I_{\Omega}=\int\int_{\Omega} \alpha A\big(\frac{K^2}{4}-\frac{2A_R}{RA^3}
-\frac{R^2A_R^2}{A^4}-4\pi(\rho-p+2p_T)\big).
\end{equation}
\textbf{Remark 2: }The quasi local mass $m$ is given by
\[
m=\frac{r}{2}(1-|\nabla r|^2).
\]
Here $r$ is the area radius, and $\nabla r$ is the gradient of $r.$
In our case $r=AR,$ which implies that
\begin{equation}\label{Hawking}
m=\frac{R^3A^3K^2}{8}-R^2A_{R}-\frac{R^3(A_{R})^2}{2A}.
\end{equation}
It thus follows that $I_{\Omega}$ can be written as
\begin{equation}\label{GC}
I_{\Omega}=\int\int_{\Omega} \alpha A\big(\frac{2m}{A^3R^3}-4\pi(\rho-p+2p_T)\big)\,dRdt.
\end{equation}
This is thus identical to the structure of the corresponding term in Schwarzschild coordinates, cf. Remark 1. Here we will however stick to the form (\ref{Iomega}) for the estimates. \\
To summarize, we have in view of (\ref{Hest}) obtained the estimate
\begin{equation}\label{Is}
H(t)\leq H(0)e^{I_{\Omega}-I_{C_t}-I_{C_{\infty}}-I_{0}},
\end{equation}
where
\begin{eqnarray}
I_{C_t}&=&-\int_{{\cal{R}}(t)}^{R_{\infty}}(KA)(t,R)\, dR,\\
I_{C_0}&=&\int_{{\cal{R}}(0)}^{R_{\infty}}(KA)(0,R)\, dR,
\end{eqnarray}
and
\begin{equation}
I_{C_{\infty}}=\int_0^t \frac{\alpha_{R}(s,R_{\infty})}{A(s,R_{\infty})}
-(\beta KA)(s,R_{\infty}))\,ds.
\end{equation}
We now invoke the a priori bounds derived in the previous section. Since ${\cal{R}}(s)\geq
\epsilon,\, 0\leq s\leq t,$ we have in view of Lemma \ref{lemma4mi},
\[
I_{\Omega}\leq \int\int_{\Omega} \alpha A\frac{2m}{A^3R^3}\,dRdt\leq \frac{2t}{\epsilon}.
\]
Using the bounds (\ref{os}), (\ref{Aestimate}) and (\ref{Kestimate}) we obtain
\[
I_{C_t}\leq \frac{C(1+t^2)}{\epsilon}.
\]
The integral $I_{C_{\infty}}$ vanishes in view of Lemma \ref{lemmadecay} and
$I_{C_0}$ depends only on the initial data. Since the second term in (\ref{Hest}) can 
be estimated in the same way we obtain 
\[
H(t)\leq (H(0)+\frac{Ct}{\epsilon^2})e^{\frac{C(1+t^2)}{\epsilon}}.
\]
The estimate for $G$ is analogous. 
Thus
\[
G(t)\leq (G(0)+\frac{Ct}{\epsilon^2})e^{\frac{C(1+t^2)}{\epsilon}},
\]
and we get
\[
Q(t)\leq (Q(0)+\frac{Ct}{\epsilon^2})e^{\frac{C(1+t^2)}{\epsilon}}.
\]
This completes the proof of Theorem \ref{theorem3}.
\begin{flushright}
$\Box$
\end{flushright}


\begin{thebibliography}{AAAA}
\bibitem{A1}
{\sc H.~Andr\'{e}asson},
On global existence for the spherically symmetric
Einstein-Vlasov system in Schwarzschild coordinates,
{\em Indiana Univ.~Math.~J.}\ {\bf 56}, 523--552 (2007).

\bibitem{A2}
{\sc H.~Andr\'{e}asson},
The Einstein-Vlasov System/Kinetic Theory,
{\em Living Rev. Relativity}\ {\bf 8}, (2005).

\bibitem{AKR1}
{\sc H.~Andr\'{e}asson, M.~Kunze, G.~Rein},
Global existence for the spherically symmetric Einstein-Vlasov
system with outgoing matter,
{\em Comm.\ Partial Differential Eqns.}\
{\bf 33}, 656--668 (2008).

\bibitem{AKR2}
{\sc H.~Andr\'{e}asson, M.~Kunze, G.~Rein},
The formation of black holes in spherically symmetric gravitational
collapse, arXiv:0706.3787.

\bibitem{AKR3}
{\sc H.~Andr\'{e}asson, M.~Kunze, G.~Rein}, Gravitational collapse and
the formation of black holes for the spherically symmetric
Einstein-Vlasov system,
{\em Quart.\ Appl.\ Math.}, {\bf 68}, 17--42 (2010).

\bibitem{AR1}
{\sc H.~Andr\'{e}asson, G.~Rein},
The asymptotic behaviour in Schwarzschild time
of Vlasov matter in spherically
symmetric gravitational collapse, {\em Math. Proc. Camb. Phil. Soc.,} {\bf 149}, 
173--188 (2010). 

\bibitem{AR2}
{\sc H.~Andr\'{e}asson, G.~Rein},
Formation of trapped surfaces for the spherically symmetric Einstein-Vlasov system,
{\em J. Hyperbolic Diff. Equations,} to appear.

\bibitem{B}{}{\sc J. Batt, }Global symmetric solutions of the initial
  value problem of stellar dynamics, {\em J. Diff. Eqns., } {\bf 25}, 342--364 (1977).

\bibitem{Cu1}{}{\sc D. Christodoulou, }On the global initial value
  problem and the issue of singularities, {\em Class. Quantum
  Grav., } {\bf 16}, A23--A35 (1999). 

\bibitem{Cu2}
{\sc D. Christodoulou, }Bounded variation solutions of the spherically symmetric Einstein-scalar field equations, {\em Comm. Pure Appl. Math., } {\bf 46}, 1131--1220 (1993).

\bibitem{Ds1}{}{\sc M. Dafermos, }Spherically symmetric spacetimes with
  a trapped surface, {\em Class. Quantum Grav., } {\bf 22}, 2221--2232 (2005).

\bibitem{Ds2}{}{\sc M. Dafermos, }
A note on the collapse of small data self-gravitating massless collisionless matter,
{\em J. Hyperbolic Diff. Eqs. }{\bf 3}, 589--598 (2006).

\bibitem{DR}
{\sc M.~Dafermos, A.~D.~Rendall},
An extension principle for the Einstein-Vlasov system in spherical symmetry,
{\em Ann.\ Henri Poincar\'{e}}\ {\bf 6}, 1137--1155 (2005).

\bibitem{LP}{}{\sc P.L. Lions and B. Perthame, }Propagation of moments
  and regularity for the 3-dimensional Vlasov-Poisson system, {\em
    Invent. Math., } {\bf 105}, 415--430 (1991).

\bibitem{MO}{\sc E. Malec and N. \'{O} Murchada},
Optical scalars and singularity avoidance in spherical spacetimes,
{\em Phys. Rev., } {\bf 50}, 6033--6036 (1994).

\bibitem{P}{}{\sc K. Pfaffelmoser, }Global classical solutions of
the Vlasov-Poisson system in three dimensions for general initial
data, {\em J. Diff. Eqns. } {\bf 95}, 281--303 (1992).

\bibitem{R}{}{\sc G. Rein, }The Vlasov-Einstein system with surface
  symmetry, {\em Habilitationsschrift, }Munich, (1995).

\bibitem{RR}{}{\sc G. Rein and A.D. Rendall, }Global existence of
  solutions of the spherically symmetric Vlasov-Einstein system with
  small initial data, {\em Commun. Math. Phys., } {\bf 150}, 561--583
  (1992). Erratum: {\em Commun. Math. Phys., } {\bf 176}, 475--478 (1996).

\bibitem{RRS}{}{\sc G. Rein, A.D. Rendall and J. Schaeffer, }
A regularity theorem for solutions of the spherically symmetric
  Vlasov-Einstein system, {\em Commun. Math. Phys. } {\bf 168}, 467--478 (1995).

\bibitem{R1}{}{\sc A.D. Rendall, }An
introduction to the Einstein-Vlasov system, {\em Banach Center Publ.,
  } {\bf 41}, 35--68 (1997).

\bibitem{R2}{}{\sc A.D. Rendall, }Cosmic censorship
and the Vlasov equation, {\em Class. Quantum Grav. } {\bf 9}, L99--L104 (1992).



\end{thebibliography}
\end{document}